\documentclass[prb,twocolumn,showpacs,floatfix,superscriptaddress]{revtex4}%
\usepackage{graphicx}
\usepackage{dcolumn}
\usepackage{bm}
\usepackage{latexsym}

\begin{document}

\title{Magnetic Moment Formation in Quantum Point Contacts}
\author{P. S. Cornaglia}
\affiliation{Centro At\'{o}mico Bariloche, 8400 San Carlos 
de Bariloche, R\'{\i}o Negro, Argentina}
\affiliation{CEA-Saclay, DSM/DRECAM/SPCSI, B\^at 462, F-91191 Gif sur Yvette, France}
\author{C.A. Balseiro}
\affiliation{Centro At\'{o}mico Bariloche, 8400 San Carlos 
de Bariloche, R\'{\i}o Negro, Argentina}
\author{M. Avignon}
\affiliation{Laboratoire d'Etudes des Propri\'et\'es Electroniques des Solides, Associated with Universit\'e Joseph Fourier, C.N.R.S, BP 166, 38042 Grenoble Cedex 9, France}

\date{today}

\begin{abstract}
We study the formation of local magnetic moments in quantum point contacts.
Using a Hubbard-like model to describe point contacts formed in a two
dimensional system, we calculate the magnetic moment using the unrestricted
Hartree approximation. We analyze different type of potentials to
define the point contact, for a simple square potential we calculate a phase
diagram in the parameter space (Coulomb repulsion - gate voltage). We also
present an analytical calculation of the susceptibility to give explicit
conditions for the occurrence of a local moment, we present a simple scaling
argument to analyze how the stability of the magnetic moment depends on the
point contact dimensions.
\end{abstract}


\maketitle
\section{Introduction}

The problem of electric charge transport through quantum point contacts
(QPC) has been intensively studied both from the theoretical and the
experimental points of view.\cite{Wees1988} The observation of conductance
quantization in a diversity of point contacts is now well established and,
in general terms, understood. The discovery of extra structure, that looks
like conductance plateaus at $0.7(2e^{2}/h)$ in GaAs devices, however, still
remains as an open question.\cite
{Thomas1996,Reilly2001,Kristensen2000,Reilly2002,Seelig2003,Graham2003,Sushkov2003,Starikov2003} The recent
accumulation of experimental evidence as well as the theoretical analysis of
this structure suggest that it is due to magnetic fluctuations. \cite
{Cronenwett2002,Cornaglia2003c,Meir2002}

The experimental data around the anomalous $0.7$ plateau observed in some of
these QPC have been interpreted as due to a Kondo effect.\cite
{Cronenwett2002} This interpretation is based on the observation of a zero
bias anomaly in the conductance, its temperature and magnetic field
dependence as well as a single energy scale $k_BT_{K}$ associated with it.
The Kondo effect is due to the magnetic screening of a localized spin, a
phenomenon that occurs for magnetic impurities diluted in metals \cite
{Hewson} and also for quantum dots in mesoscopic circuits.\cite{Gordon1998}
While all the $0.7$ anomaly features observed in QPC are consistent with the
occurrence of Kondo effect, it is not clear how a magnetic moment could
develop in these systems. Point contacts in GaAs-AlGaAs heterostructures are
built as narrow constrictions in a two dimensional electron gas formed at
the interface between GaAs and AlGaAs. The constriction is built using
patterned surface depletion gates so the shape and size of the QPC can be
controlled.

It has been shown that a two dimensional electron gas with a constriction
under some special conditions, that can be achieved by tuning a gate
voltage, can develop a magnetic moment. This has been done by using spin
dependent density functional theory\cite{Hirose2003} on the one hand and
unrestricted Hartree-Fock solutions of effective Hubbard models on the
other. \cite{Cornaglia2003c} As already pointed out, one should be careful
in the physical interpretation of the frozen spin solution obtained with
these methods. As in the old impurity problem of magnetic moment formation, 
\cite{Anderson1961} the mean field solution can not be correct since it
breaks the local symmetry. In the exact solution of the problem spin
fluctuations should recover local rotational invariance. It is very
difficult to incorporate spin fluctuations on top of the mean field solution
to fully recover the rotational invariance. However the mean field solution
gives a good indication of the region in parameter space where we should
expect magnetic fluctuations to play a central role in the low energy
physics.\cite{Hewson} What is still missing in the problem of the magnetic
nature of QPC is a detailed analysis of the condition for the occurrence of
a magnetic moment including its geometry and size dependence. In this work
we use the Hartree criteria to determine the region of the parameter space
where the contact develops a magnetic moment. In next section we present the
model and its mean-field version. In section \ref{sub:Num}, the numerical
solution is used to determine the region of stability of a localized
magnetic moment in the point contact. In part \ref{sub:Ana} the analytical
expressions for the mean field magnetic instability are presented, and in
part \ref{sub:Sca} the limit of narrow resonances is used to predict size
scaling. The last section includes summary and discussion.

\section{The Magnetic Instability at the Point Contact}

A natural description of the two dimensional (2D) electron gas in
GaAs-AlGaAs heterostructures is an effective mass theory in which the
kinetic energy is given by the one of a 2D Fermi gas of particles with an
effective mass $m^{\ast }$ and a characteristic particle density $n\sim
10^{11}/cm^{2}$. For this system the Fermi wave vector $k_{F}$ is of the
order of $10^{6}/cm$. The potential created by the applying gate voltages is
described by a function $V(\mathbf{r})$ where $\mathbf{r}$ is the 2D
coordinate. Finally, to describe the magnetic properties of the system, the
electron - electron interaction has to be included explicitly. In order to
solve the Schr\"{o}dinger equation in the Hartree approximation, we
discretize the space to end up with an effective tight binding model with
hopping matrix element $t=$ $\hbar ^{2}/2m^{\ast }a^{2}$ where $a$ is the
discretization parameter, the lattice parameter of the tight binding model.
The potential energy $V(\mathbf{r})$ is included as an on site energy and
the Coulomb repulsion as an on site Hubbard term. The discretization
parameter $a$ should be taken such that the Fermi wave vector $k_{F}$ is
much smaller than any reciprocal lattice vector, typically $k_{F}\ll 2\pi /a$%
.

Our starting point is a two dimensional extended Hubbard Hamiltonian\cite
{yeyati-flores} with a constriction and a gate potential, $H=H_{0}+H_{int}$
with

\begin{equation}
H_{0}=\sum_{i,\sigma }\varepsilon _{i}c_{i\sigma }^{+}c_{i\sigma
}-t\sum_{\langle ij\rangle ,\sigma }c_{i\sigma }^{+}c_{j\sigma }  \label{ham}
\end{equation}
and 
\begin{equation}
H_{int}=U[\sum_{i}c_{i\uparrow }^{+}c_{i\uparrow }c_{i\downarrow
}^{+}c_{i\downarrow }+\frac{1}{2}\sum_{i\neq j,\sigma ,\sigma 
{\acute{}}}\eta _{ij}c_{i\sigma }^{+}c_{i\sigma }c_{j\sigma {\acute{}}}^{+}c_{j\sigma {\acute{}}
}].
\end{equation}
Here $c_{i\sigma }^{+}$ creates an electron at site $i$ with spin $\sigma $,
the diagonal energy $\varepsilon _{i}$ is a function of the coordinate $\mathbf{r}_{i}$ of site $i$ and defines the point contact with the gate
potential. The last term in $H_{0}$ the the site independent nearest
neighbor hopping. The parameter $U$ is the local Coulomb repulsion and the
dimentionless parameter $\eta _{ij}<1$ gives the spacial dependence of a
screened interaction.

In the Hartree approximation the potential energy is renormalized by the
Coulomb repulsion
\begin{equation}
H_{MF}=\sum_{i,\sigma }\widetilde{\varepsilon }_{i\sigma }c_{i\sigma
}^{+}c_{i\sigma }-t\sum_{\langle ij\rangle ,\sigma }c_{i\sigma
}^{+}c_{j\sigma }-K,  \label{ham-hf}
\end{equation}
with $\widetilde{\varepsilon }_{i\sigma }=\varepsilon _{i}+U[\langle
n_{i,-\sigma }\rangle +\sum_{j,\sigma 
{\acute{}}}(1-\delta _{ij})\eta _{ij}\langle n_{j,\sigma  {\acute{}}}\rangle ]$, $n_{i,\sigma }=c_{i\sigma }^{+}c_{i\sigma }$ is the number
operator and $\langle ..\rangle $ indicates the expectation value and $K$ is
a constant. In what follows we use the notation 
\begin{eqnarray}
n_{i} &=&\langle n_{i,\uparrow }\rangle +\langle n_{i,\downarrow }\rangle 
\nonumber \\
m_{i} &=&(\langle n_{i,\uparrow }\rangle -\langle n_{i,\downarrow }\rangle
)/2  \label{nn}
\end{eqnarray}
so that $\widetilde{\varepsilon }_{i\sigma }=\widetilde{\varepsilon }%
_{i}-\sigma Um_{i}$ with $\widetilde{\varepsilon }_{i}=\varepsilon
_{i}+U[n_{i}/2+\sum_{j}(1-\delta _{ij})\eta _{ij}n_{j}]$.

\begin{figure}[tbp]
\includegraphics[width=8.0cm,clip=true]{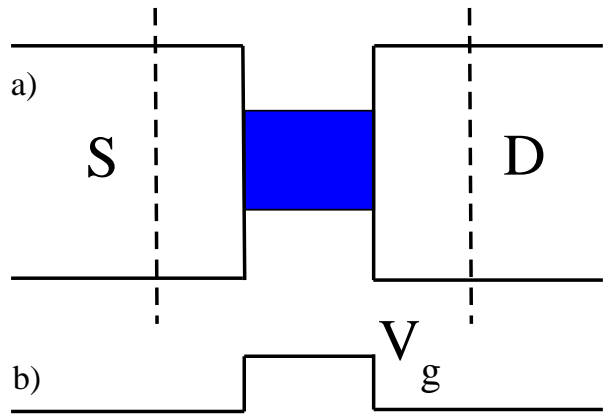}
\caption{a) Schematic picture of the model for the QPC (dark region) between
the source and drain leads, the potential at the constriction is $V_{g}$. b)
Profile of the potential along the QPC. }
\label{fig:fig1}
\end{figure}

The shape of the QPC as well as the gate voltage is defined through the
potential energy $\varepsilon _{i}$. We study systems with the geometry of
Fig. \ref{fig:fig1} consisting of a two dimensional strip infinitely long
and with a width of $N$ sites. The QPC is defined as a constriction at the
center of the strip. In the rest of the work, we assume a square lattice and
weak Coulomb repulsion $U$, much smaller than the Stoner critical value. We
also take the average charge per site $n_{0}<1$ to avoid any Fermi surface
nesting effect. Then far from the QPC the charge is uniform and the
magnetization is zero so that the potential energy is $\widetilde{%
\varepsilon }_{i\sigma }=Un_{0}[1/2+\sum_{j}(1-\delta _{ij})\eta _{ij}]$.
To do the calculation, we artificially divide the sample in two regions: a
central region (between vertical dashed lines in Fig. \ref{fig:fig1}) that
includes the QPC and the uniform regions that include the source and drain
far away from the point contact. In the numerical calculation we evaluate
the self-consistent solution within the unrestricted Hartree scheme for the
central region coupled with the lateral regions with uniform charge and zero
magnetization. This procedure is acceptable if the charge and magnetization
profiles are continuous and smooth at the boundary between regions. The same
scheme is used to analyze the spin susceptibility. 

\subsection{Numerical Solution of the Unrestricted Hartree Equations}

\label{sub:Num}

In this section we present the numerical solution of the unrestricted
Hartree approximation. The charge $n_{i}$ and magnetization $m_{i}$ at each
site of the central region containing the QPC are evaluated with the
solution of the self-consistent equations: 
\begin{equation}
\langle n_{i,\sigma }\rangle =-\frac{1}{\pi }\int^{\varepsilon _{F}}d\omega {%
Im}\left[ G_{i,\sigma }^{R}(\omega )\right] 
\end{equation}
here the retarded Green function $G_{i,\sigma }^{R}(\omega )$ is the
diagonal element of 
\begin{equation}
G^{R}=\left[ \omega +i0^{+}-H_{MF}^{0}-\Sigma ^{R}(\omega +i0^{+})\right]
^{-1}
\end{equation}
where $H_{MF}^{0}$ is the Hartree Hamiltonian of the central part and $%
\Sigma ^{R}(\omega +i0^{+})$ is the self energy due to the lateral regions
with $m_{i}=0$.

\begin{figure}[tbp]
\includegraphics[width=8.0cm,clip=true]{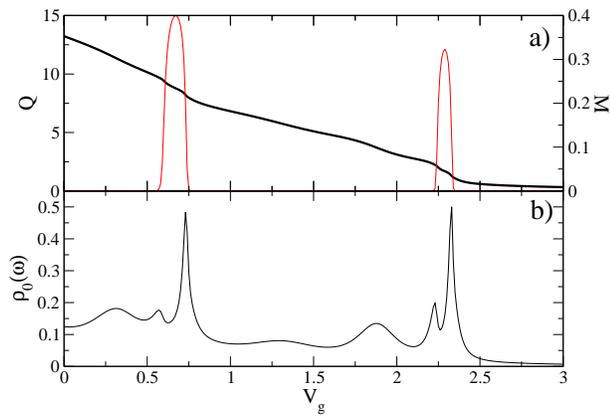}
\caption{(a) Magnetic moment (thin line) and total charge (thick line) at
the constriction of a $3$-site-wide and $9$-site-long QPC. The local
interaction is $U=2.0$ and the Fermi energy is $\protect\epsilon_F=-1$, with
all parameters in units of $t=1$. The source and drain slabs have a width of 
$20$ sites. b) Local density of states at the Fermi level, averaged over the
sites of the constriction, calculated in the restricted Hartree
approximation. }
\label{fig:fig2}
\end{figure}

\begin{figure}[tbp]
\begin{center}
\includegraphics[width=8cm,clip=true]{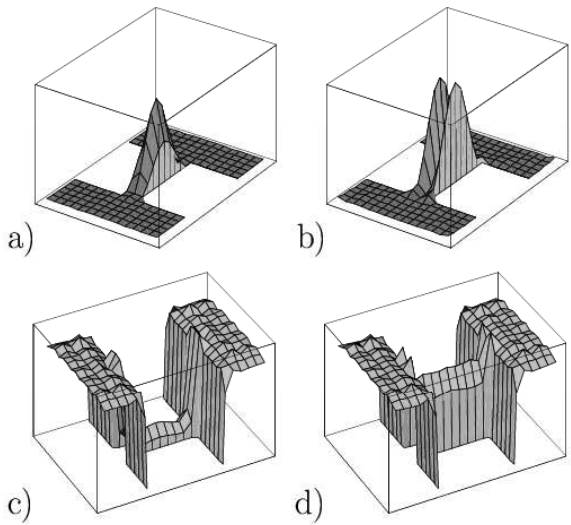}
\end{center}
\caption{Spatial distribution of the a) magnetization, c) charge for $%
V_{g}=0.66$, and other parameters as in Fig. \ref{fig:fig2}. b) and d) Same
as a) and c) respectively with $V_{g}=2.27$.}
\label{fig:dens}
\end{figure}
We studied different  QPC shapes. In the present approximation, the long
range part of the electron-electron interaction renormalizes the on site
energy and redefines the shape of the QPC. From hereon we take $\eta _{ij}=0$%
. We first present results for a model square potential shown in Fig. \ref
{fig:fig1} and defined as: $\varepsilon _{i}=0$, $V_{g}$ and $\infty $ for $i
$ in the source and drain, in the neck of the QPC and at the sides of it
respectively. We considered a variety of QPC with $N_{0}=3$ sites in the
lateral direction and $N_{1}=9$ sites in the longitudinal direction. We
define the total charge and the magnetization of the contact as $%
Q=\sum_{i}^{^{\prime }}n_{i}$ and $M=\sum_{i}^{^{\prime }}m_{i}$ where the
sum is over all the sites of the QPC. The self-consistent results are shown
in Fig. \ref{fig:fig2} for a QPC of width $N_{0}$ and length $N_{1}$. For
large gate voltages the total charge $Q$ in the point contact is
exponentially small. As $V_{g}$ decreases $Q$ increases and for some values
of the gate potential there is an abrupt increase in the total charge. The
steps obtained around this point correspond to approximately one electron
been transferred from the source and drain to the point contact. This
behavior in $Q$ is characteristic of Coulomb blockade. Between the two first
steps in $Q$ there is a spin 1/2 localized at the point contact as indicated
by the magnetization curve versus $V_{g}$ shown in Fig. \ref{fig:fig2}(a).
The local density of states at the QPC shown in Fig. \ref{fig:fig2}(b)
presents a series of resonances. The resonances are associated with
longitudinal modes in the QPC. The occurrence of a local magnetization at
the QPC coincides with a narrow resonance crossing the Fermi energy. The
first resonance has no nodes in the QPC as shown in the magnetization
profile of Fig. \ref{fig:dens}(a). As $V_{g}$ decreases other resonances
cross the Fermi energy and new magnetic solutions are obtained. When the
first resonance of the second channel cross the Fermi energy, again a spin $%
1/2$ is localized in the QPC. The wave functions of the second channel have
a node at the center of the QPC in the transverse direction and generate the
magnetization profile shown in Fig. \ref{fig:dens}(b). For the value of $U$
considered in Fig. \ref{fig:fig2} magnetic moments are obtained only for the
first resonance of each transverse channels. However, the other wider
resonances may also produce local magnetic moments for larger values of $U$. 
\begin{figure}[tbp]
\begin{center}
\includegraphics[width=4cm,clip=true]{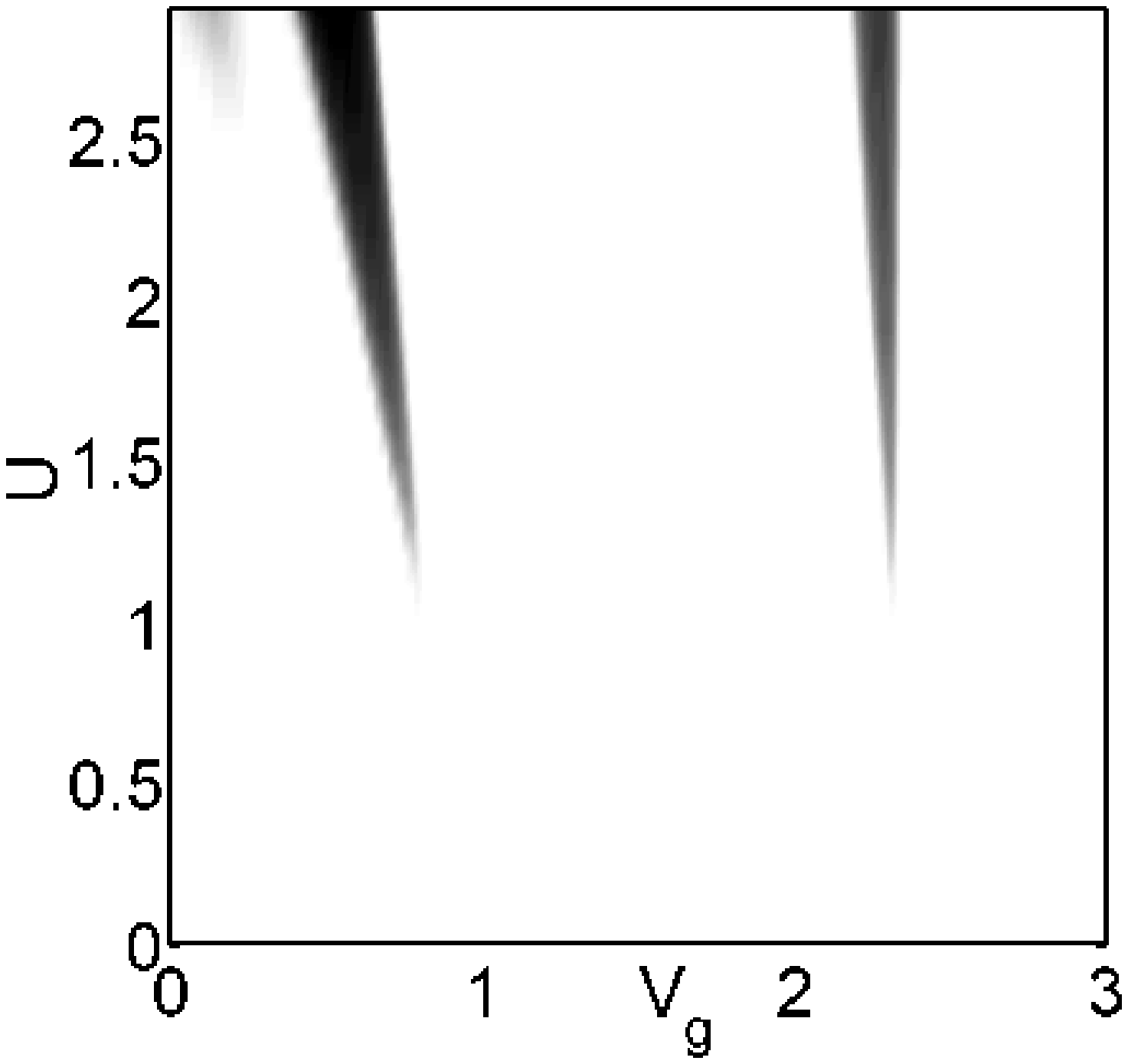} %
\includegraphics[width=4cm,clip=true]{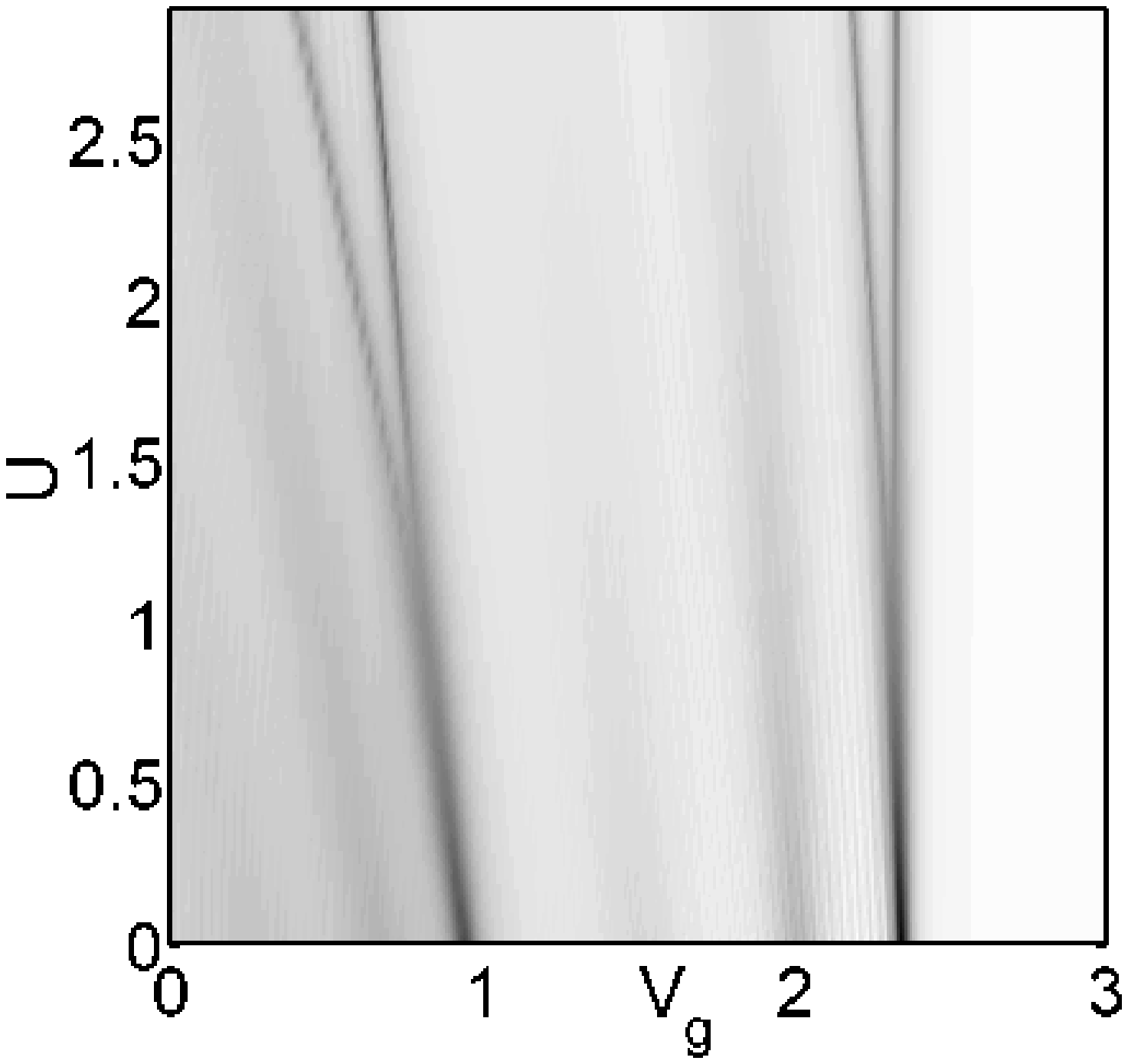}
\end{center}
\caption{Phase diagram $U$--$V_{g}$ for a rectangular $3\times 9$ QPC, dark
regions indicate higher values. Left: magnetization at the QPC, Right:
negative derivative of the charge with respect to $V_{g}$.}
\label{fig:PD3x9}
\end{figure}
In Fig. \ref{fig:PD3x9}(a) we present the phase diagram in the parameter
space [$U-$ $V_{g}$] for the $3\times 9$ QPC presented up to now. The dark
regions correspond to a stable magnetic moment at the QPC. Fig. \ref
{fig:PD3x9}(b) illustrates the behavior of the charge $Q$; to analyze the
behavior of $Q$ it is convenient to plot its derivative $-\partial
Q/\partial V_{g}$. For small values of $U$ we obtain non-magnetic solutions
for all values of $V_{g}$. However as $V_{g}$ increases and the QPC
resonances cross the Fermi level, $Q$ increases and there is a maximum in $%
-\partial Q/\partial V_{g}$. For the parameters of the figure, magnetic
solutions are obtained for $U/t\gtrsim 1$ and for values of the gate
potential $V_{g}$ that make the resonances to coincide with the Fermi level.
The occurrence of a magnetic solution is accompanied by a splitting of the $%
(-\partial Q/\partial V_{g})$ maximum, a characteristic of the Coulomb
blockade regime. Fig. \ref{fig:PD5x6} shows the phase diagram and the
behavior of $Q$ for a shorter an wider point contact. In this case larger
values of U are required to produce magnetic moments. In next section we
interpret these results in terms of the susceptibility evaluated in a simple
approximation and make a scaling analysis to describe how the magnetic
instability depends on the QPC size. 
\begin{figure}[tbp]
\begin{center}
\includegraphics[height=4cm,clip=true]{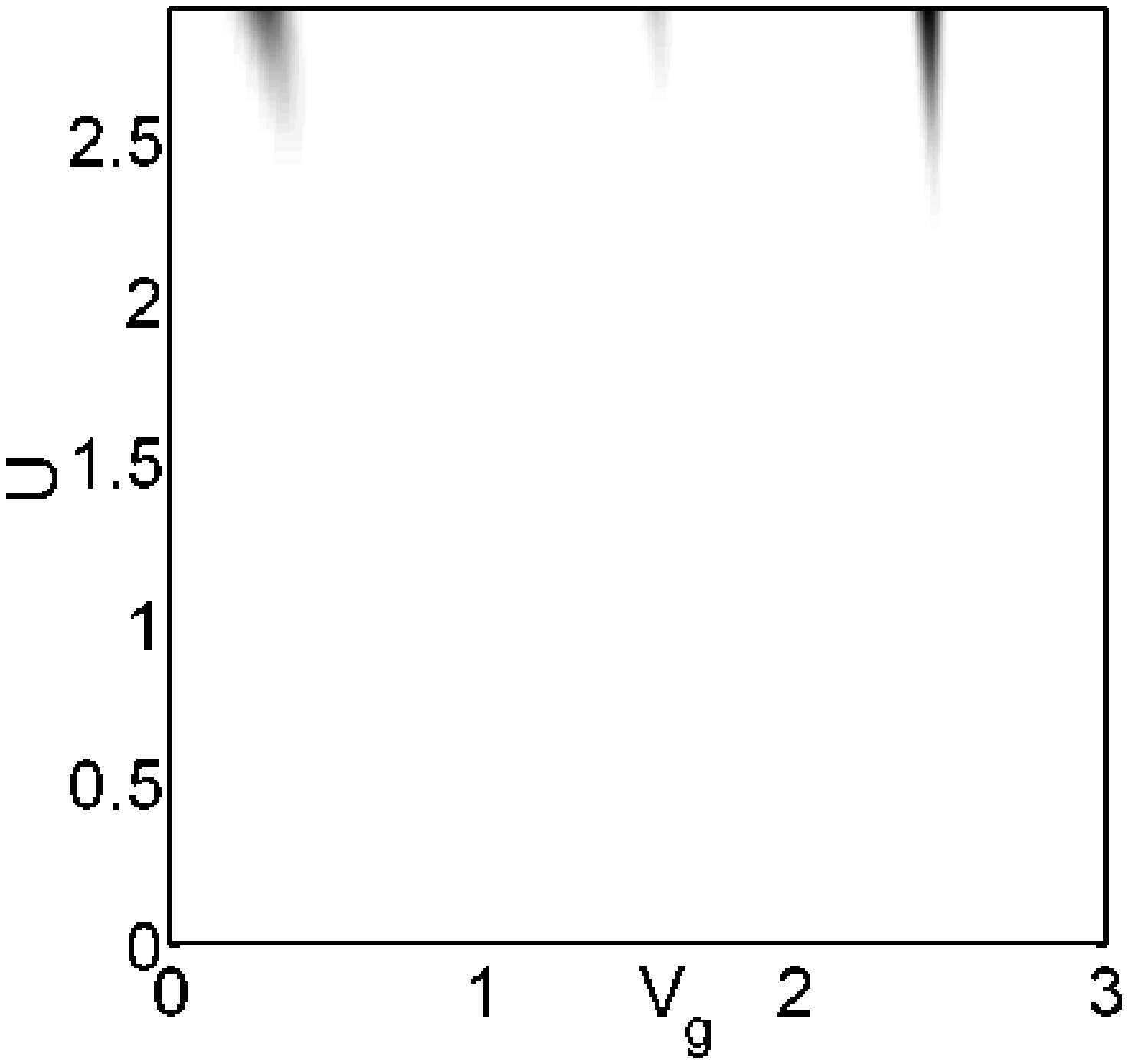} %
\includegraphics[height=4cm,clip=true]{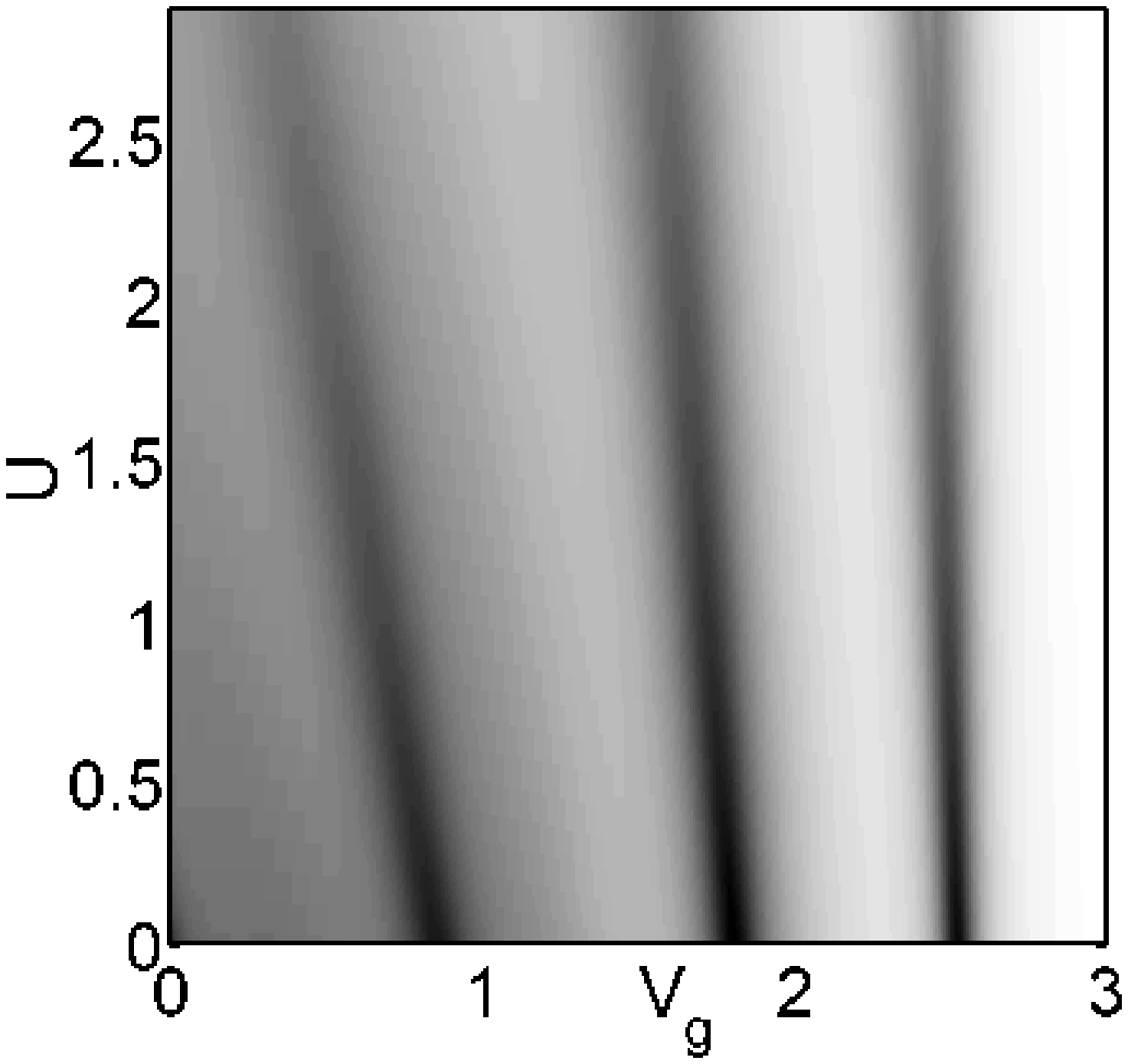}
\end{center}
\caption{Phase diagram $U$--$V_{g}$ for a rectangular $5\times 6$ QPC, dark
regions indicate higher values. Left: magnetization at the QPC, Right:
negative derivative of the charge with respect to $V_{g}$.}
\label{fig:PD5x6}
\end{figure}
The square potential optimizes resonances at the QPC and consequently favors
the formation of local moments. We end this section showing some results
obtained with a smoother and more realistic potential defined as: 
\begin{eqnarray}
V(\mathbf{r}_{i}) &\equiv &\varepsilon
_{i}=V_{g}f(y_{i}-a_{1}/2)[1-f(y_{i}+a_{1}/2)]\times   \nonumber \\
&&\left\{ 1+10\left[ f(x_{i}+a_{0}/2)+f(-x_{i}+a_{0}/2)\right] \right\} ,
\end{eqnarray}
where $\mathbf{r}_{i}=(x_{i},y_{i})$ is the coordinate of site $i$, $%
f(x)=(1+e^{x/\lambda })^{-1}$with $\lambda $ a characteristic screening
length, $a_{0}$ and $a_{1}$ are the QPC width and length respectively. To
describe a wedge-like point contact we take a width that varies linearly
with $y_{i}$: $a_{0}(y_{i})=a_{0}+\delta a_{0}\times (y_{i}+a_{1}/2)/a_{1}$.
The potential profiles for these contacts are shown in Fig. \ref{fig6}. In
the same figure the average density of states at the point contact is also
shown. For a wire-like point contact resonant states are obtained, the
energies of these resonances are at the bottom of each channel. The density
of states has much less structure in the case of wedge-like contacts. This
shows that wire-like contacts are good candidates to develop magnetic
fluctuations while wedge-like structures are not. 
\begin{figure}[tbp]
\begin{center}
\includegraphics[width=8cm,clip=true]{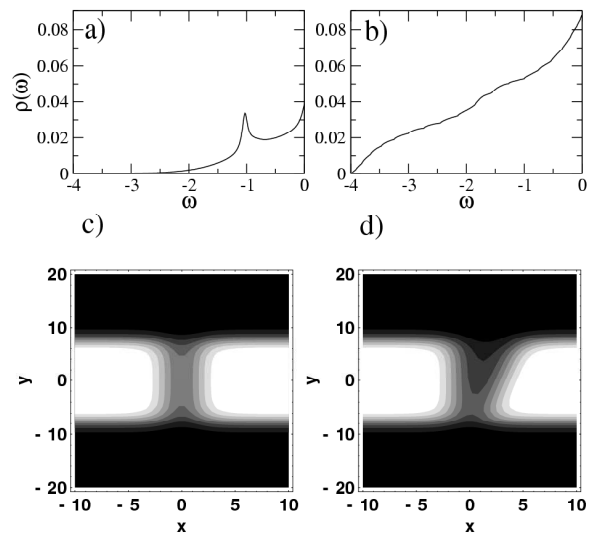}
\end{center}
\caption{a) Average local density of states in the QPC, at the Fermi level
for the rectangular smooth potential of c); $U=0$, $V_{g}=0.5$, $\protect%
\varepsilon _{F}=-1$, $a_{0}=3$, $a_{1}=15$. b) Same as a) for the wedge
shaped potential of d) and $V_{g}=1.25$, $\protect\delta a_{0}=4.5$.}
\label{fig6}
\end{figure}

\subsection{Spin Susceptibility}

\label{sub:Ana} In the presence of an external magnetic field, the local
magnetization is given by 
\begin{equation}
m_{i}=\sum_{j}\chi _{i,j}h_{j}  \label{mxh}
\end{equation}
where $h_{j}$ is the magnetic field at site $j$ and $\chi _{i,j}$ is the
non-local susceptibility. In our system with no translational symmetry the
non-local susceptibility depends on the two coordinates$\ \mathbf{r}_{i}$
and $\mathbf{r}_{j}$. In the Hartree approximation, the energy shift of
the one particle levels is used to define an effective magnetic field given
by $h_{j}=h_{ext}+Um_{j}$ where we assume a \ uniform external field $%
h_{ext} $. The magnetization is then given by 
\begin{equation}
m_{i}=U\sum_{j}\chi _{i,j}m_{j}+\overline{\chi }_{i}h_{ext}  \label{mi}
\end{equation}
with $\overline{\chi }_{i}=\sum_{j}\chi _{i,j}$. As we show below, $%
\overline{\chi }_{i}$ is just the local density of states at the Fermi
energy and for $U=0$ the Pauli susceptibility depends on the coordinate as
the local density of states. For $h_{ext}=0$ a non-trivial solution of
equation (\ref{mi}) gives the onset of a spontaneous magnetization. We
define the susceptibility matrix $\mathbf{\chi}$, with matrix elements $\chi
_{i,j}$, and a magnetization vector $\mathbf{M}$\ as a column vector with
components $m_{i}$. Then equation (\ref{mi}) for $h_{ext}=0$ has the form of
an eigenvalue problem 
\begin{equation}
\mathbf{\chi M=}\frac{1}{U}\mathbf{M}  \label{auto}
\end{equation}
\ as $U\rightarrow 0$ there is no non-trivial solution of this equation. The
onset of a spontaneous magnetization is given by the largest eigenvalue of $%
\mathbf{\chi }$ being equal to $1/U$ , the corresponding eigenvector gives
the magnetic profile of the instability.

Since in the paramagnetic state the system has spin rotational invariance,
we calculate the transverse susceptibility. The linear response of the
system to a magnetic field along the $\mathbf{x}$ direction is: 
\begin{equation}
\chi _{i,j}=-\ll c_{i\uparrow }^{+}c_{i\downarrow },c_{j\downarrow
}^{+}c_{j\uparrow }\gg _{\omega =0}
\end{equation}
where $\ll A,B\gg $ indicates the retarded Green function. In terms of the
self-consistent one-particle eigenstates of the Hamiltonian $H_{HF}$ with
wave functions $\varphi _{\nu }(i)$ and energies $\varepsilon _{\nu }$, the
susceptibility is given by

\begin{equation}
\chi _{i,j}=-\sum_{\nu \mu }\varphi _{\nu }^{*}(i)\varphi _{\mu }(i)\varphi
_{\mu }^{*}(j)\varphi _{\nu }(j)\frac{f(\varepsilon _{\nu })-f(\varepsilon
_{\mu })}{\varepsilon _{\nu }-\varepsilon _{\mu }}  \label{xi}
\end{equation}
where $f(\varepsilon )$ is the Fermi function.

Due to the orthogonality of the one-particle wave functions, at zero
temperature, we have: 
\begin{eqnarray}
\overline{\chi }_{i} &=&\sum_{j}\chi _{i,j}=-\sum_{\nu }|\varphi _{\nu
}(i)|^{2}\frac{\partial f(\varepsilon _{\nu })}{\partial \varepsilon _{\nu }}
\\
&=&\sum_{\nu }|\varphi _{\nu }(i)|^{2}\delta (\varepsilon _{\nu
}-\varepsilon _{F})=\rho _{i}(\varepsilon _{F})  \nonumber
\end{eqnarray}
where we have taken the ratio between the Fermi function difference and the
energy deference as the Fermi function derivative when $\varepsilon _{\nu
}\rightarrow \varepsilon _{\mu}$, $\rho _{i}(\varepsilon )$ is the local
density of states and $\varepsilon _{F}$ the Fermi energy. Then, inserting
the above expression in equation (\ref{mi}), for $U=0$ we obtain $m_{i}=\rho
_{i}(\varepsilon _{F})h_{ext}$.

For $U\neq 0$ in a system with no translational invariance we have to
calculate the non-local susceptibility, a matrix of dimension equal to the
number of sites of the sample. To simplify the problem we assume that $U$ is
much smaller than the Stoner value that generates a global instability,
consequently a non trivial solution of equation (\ref{auto}) should have the
magnetization concentrated in the region of the point contact. We assume
that far from the contact $m_{i}=0$ and look for solutions with $m_{i}\neq 0$
only if $i$ belongs to a small region $\ R$ that includes the contact. Then
we have to calculate a reduced matrix susceptibility $\chi _{i,j}$ with $i,j$
in $R$.

Since even in the paramagnetic state, and due to the charge redistribution
close to the point contact, the one-particle states of $H_{HF}$ depend on $U$
in a non-trivial way, the largest eigenvalue $\kappa $ of the susceptibility
depends on $U$ and the instability condition is a self-consistent equation: 
\[
\kappa (U)=1/U 
\]
and its solutions have to be obtained numerically.

\subsection{The Resonant State Approximation}

\label{sub:Sca}

Here we present some analytical results based on the fact that, for high
gate voltages, the local density of states at the point contact present well
defined resonances. By varying the gate voltage the position of the
resonances can be tuned to coincide with the Fermi level. When this occurs,
for small quantum point contacts where the quantization effects are
important, the transport and magnetic properties of the contact are
dominated by a single resonant state and in what follows we consider this
situation. This is a valid approximation as long as the width of the
resonance remains much smaller that the separation between resonances. The
non-local susceptibility is now given by 
\begin{equation}
\chi _{i,j}=-\widetilde{\sum_{\nu \mu }}\varphi _{\nu }^{*}(i)\varphi _{\mu
}(i)\varphi _{\mu }^{*}(j)\varphi _{\nu }(j)\frac{f(\varepsilon _{\nu
})-f(\varepsilon _{\mu })}{\varepsilon _{\nu }-\varepsilon _{\mu }}
\end{equation}
here the decorated sum indicates summation over all states belonging to a
single resonance. For $i$ in the point contact, the corresponding
wavefunctions can be written as

\begin{equation}
|\varphi _{\nu }(i)|^{2}=|\alpha _{m}(i)|^{2}\frac{\gamma /\pi \rho _{\nu }}{%
(\varepsilon _{\nu }-\Delta )^{2}+\gamma ^{2}}
\end{equation}
where $\alpha _{m}(i)$ is the wave function of the QPC, $\gamma $ and $%
\Delta $ are the width and the energy of the resonant state $m$ and $\rho
_{\nu }$ is the density of states. Since we can work with real
wavefunctions, they can be taken as the square root of the above expression
and the susceptibility can be put as 
\begin{equation}
\chi _{i,j}=|\alpha _{m}(i)|^{2}|\alpha _{m}(j)|^{2}\chi _{res}
\end{equation}
with the susceptibility of a resonant state given by 
\begin{eqnarray}
\chi _{res}&=&-\int d\varepsilon _{\nu } d\varepsilon _{\mu }\frac{(\gamma
/\pi )^{2}}{[(\varepsilon _{\nu }-\Delta )^{2}+\gamma ^{2}][(\varepsilon
_{\mu }-\Delta )^{2}+\gamma ^{2}]}\times  \nonumber \\
&&\frac{f(\varepsilon _{\nu })-f(\varepsilon _{\mu })}{(\varepsilon _{\nu
}-\varepsilon _{\mu })}
\end{eqnarray}

The susceptibility matrix has the form $\mathbf{\chi }=\chi _{res}\mathbf{A}$
where the matrix $\mathbf{A}$ can be put as 
\begin{equation}
\mathbf{A=}\left[ 
\begin{array}{c}
|\alpha _{m}(1)|^{2} \\ 
|\alpha _{m}(2)|^{2} \\ 
: \\ 
|\alpha _{m}(N)|^{2}
\end{array}
\right] \left[ |\alpha _{m}(1)|^{2},|\alpha _{m}(2)|^{2}..|\alpha
_{m}(N)|^{2}\right]
\end{equation}
and the instability condition becomes 
\begin{equation}
\alpha U\chi _{res}=1
\end{equation}
where $\alpha $ is the largest eigenvalue of the matrix $\mathbf{A}$. From
the form of $\mathbf{A}$ it is clear that the vector 
\begin{equation}
\mathbf{M}=\left[ 
\begin{array}{c}
|\alpha _{m}(1)|^{2} \\ 
|\alpha _{m}(2)|^{2} \\ 
: \\ 
|\alpha _{m}(N)|^{2}
\end{array}
\right]
\end{equation}
is an eigenvector with eigenvalue $\alpha =\sum_{i}|\alpha _{m}(i)|^{4}$ and
that all other eigenvalues are zero. The condition for the formation of a
magnetic moment at the point contact is

\begin{equation}
U_{eff}\chi _{res}=1
\end{equation}
where $U_{eff}=\sum_{i}|\alpha _{m}(i)|^{4}U$ is the effective Coulomb
repulsion for two electrons at the QPC state $m$ with spacial wavefunction $%
\alpha _{m}(i)$. If the resonance is centered at the Fermi level, this
condition is simply 
\begin{equation}
U_{eff}=\pi \gamma /2  \label{cond}
\end{equation}

For small $U$ and an arbitrary potential form of the point contact defined
by the potential $V(\mathbf{r}_{i})$, the Hartree solution can be used
to estimate $\alpha _{m}(i)$ and $\gamma $.

Now we compare this criterion with the full unrestricted Hartree
calculation for the lowest energy resonance. For the square potential of
Fig. \ref{fig:fig1} the $U=0$ eigenfunctions $\alpha _{0}(i)$ are: 
\begin{eqnarray}
\alpha _{0}(i)&=&\sqrt{\frac{2}{(N_{0}+1)}}\sin (\frac{\pi }{a(N_{0}+1)}%
x_{i})\times  \nonumber \\
&&\sqrt{\frac{2}{(N_{1}+1)}}\sin (\frac{\pi }{a(N_{1}+1)}y_{i})
\end{eqnarray}
here $a$ is the lattice parameter, $aN_{0}$ and $aN_{1}$ are the QPC width
and length respectively. As this wavefunction is hybridized with the right
and left reservoirs it acquires a width $\gamma =2\pi \rho V_{eff}^{2}$
where $\rho $ is the density of states of the reservoir and the effective
hybridization is 
\begin{equation}
V_{eff}\simeq t\sum_{i\in edge}|\alpha _{0}(i)|  \label{hyb}
\end{equation}
the sum is over all the sites of the QPC that are at one edge (right or
left), hybridized with the reservoir. With this estimation and the condition
of equation (\ref{cond}) we obtain the critical value $U_{c}^{*}$ of the
Coulomb repulsion for the occurrence of a magnetic solution shown in table
1. The comparison with the values obtained using the fully unrestricted
Hartree approximation is very good in particular for long and narrow
point contacts.

\begin{table}[tbp]
\caption{Critical values of $U$ for the appearance of magnetic solutions.}
\label{table1}%
\begin{ruledtabular}
\begin{tabular}{ccc}
$N_{0}\times N_{1}$ & $U_{c}/t$ & $U_{c}^{*}/t$ \\ \hline
$3\times 11$ & $0.55$ & $0.57$ \\ 
$3\times 9$ & $0.85$ & $0.84$ \\
$4\times 7$ & $1.50$ & $1.76$ \\ 
$5\times 6$ & $2.25$ & $2.49$ \\
$5\times 5$ & $3.18$ & $3.96$ \\ 
\end{tabular}
\end{ruledtabular}
\end{table}

For the second longitudinal resonance of the first channel in the case of
the $3\times 9$ geometry, we find the same $U_{eff}$ as for the first
resonance while the width of the resonance becomes $\gamma ^{\prime }\sim
3.6\gamma $ [as observed in Fig. \ref{fig:fig2}(b)], therefore giving $%
U_{c}^{\prime }/t\sim 3.1$ in good agreement with the corresponding phase
diagram shown in Fig. \ref{fig:PD3x9}.

Finally we can make a approximate scaling analysis for large systems: the
effective repulsion in a single resonance $U_{eff}=\sum_{i}|\alpha
_{m}(i)|^{4}U\sim U/(N_{0}\times N_{1})$ and the effective width of the
resonance is proportional to the square of the hybridization of equation (%
\ref{hyb}), $\gamma $ $\sim $ $\rho t^{2}/N_{1}^{3}$. These estimations lead
to a critical value of $U$ that at resonance scales with the size of the
point contact as $U_{c}$ $\sim $ $\rho t^{2}N_{0}/N_{1}^{2}$. Numerical
estimations of the mean field critical value for large systems are in
agreement with this scaling.

\section{Conclusions}

We have presented results for the formation of local magnetic moments in
point contacts. We used a Hubbard-like model to describe point contacts
formed in a two dimensional system. The contact is defined in terms of a
potential $V(\mathbf{r})$ that can be varied with a single parameter $V_{g}$
representing a gate voltage. We calculate the magnetic moment using the
unrestricted Hartree approximation. For a square potential, the system shows
a marked tendency to form a localized moment at the point contact each time
a new channel is tuned to the Fermi energy. In this conditions the critical
value of the local repulsion $U$ is almost an order of magnitude smaller
than the Stoner critical value for an homogeneous system. For the parameters
of figure \ref{fig:PD3x9} the critical value for the first resonances is $U\sim t$. Using
the effective mass and electron density characteristic of GaAs-AlGaAs
hetherostructures, we can take a lattice parameter $a=10nm$. With these
numbers, the results of the figure correspond to a point contact of $%
30nm\times 90nm$ and the critical value of $U$ gives an effective
interaction $U_{eff}\sim 0.5meV$.

In long contacts defined with a square potential, the second longitudinal
resonance may also generate a local moment for moderate values of $U$. This
is a consequence of the square potential that optimizes resonances each time
the Fermi wavelength is commensurate with the contact length. For more
realistic potentials only the first longitudinal resonance of each channel
may generate a local moment. Moreover, for wedge-like contacts we found no
evidence of moment formation in the Hubbard type models. The numerical
results are interpreted in terms of a simple one-resonance approximation. We
also present a simple a scaling argument to interpret the general dependence
of the magnetic instability with the point contact dimensions.

We end by stressing that the Hartree calculation, that breaks the spin
symmetry, only gives a criterion that allows to identify the region of
parameter space were the low temperature physics may be dominated by
magnetic fluctuations. In this particular regions a Kondo-like model may be
used to describe the spin fluctuations \cite{Cornaglia2003c,Meir2002}.

\acknowledgments
We acknowledge partial support from the International Collaboration Program
between France and Argentine CNRS-PICS 1490, Fundaci\'on Antorchas, the
CONICET and ANPCYT, grants N. 02151 and 99 3-6343. One of us (CAB) is also
grateful to Universit\'e Joseph Fourier for a visiting professorship during
which part of this work has been done.

\bibliography{referencias}

\end{document}